\documentclass[doublecol,linenumbers]{epl2}
\usepackage{amsmath}
\usepackage{amssymb}
\usepackage{graphicx}
\usepackage{psfrag}
\usepackage[normalem]{ulem}
\usepackage[utf8]{inputenc}

\title{Pairing in a system of a few attractive fermions in a harmonic trap}

\author{Tomasz Sowi\'nski\inst{1,2} \and Mariusz Gajda\inst{1,2} \and Kazimierz Rz\c a\.zewski\inst{2}}
\shortauthor{T. Sowi\'nski, M. Gajda, K. Rz\c a\.zewski}

\institute{                    
  \inst{1} Institute of Physics of the Polish Academy of Sciences, Al. Lotnik\'ow 32/46, 02-668 Warszawa, Poland\\
  \inst{2} Center for Theoretical Physics of the Polish Academy of Sciences, Al. Lotnik\'ow 32/46, 02-668 Warszawa, Poland
  }

\abstract{
We study a strongly attractive system of a few spin-1/2 fermions confined in a one-dimensional harmonic trap, interacting via two-body contact potential. Performing exact diagonalization of the Hamiltonian we analyze the ground state and the thermal state of the system in terms of  one-- and two--particle reduced density matrices. We show how  for strong attraction the correlated pairs emerge in the system. We find that the fraction of correlated pairs depends on temperature and we show that this dependence has universal properties analogous to the gap function known from the theory of superconductivity. In contrast to the standard approach based on the variational ansatz and/or perturbation theory, our predictions are exact and are valid also in a strong attraction limit.  Our findings contribute to the understanding of strongly correlated few-body systems and can be verified in current experiments on ultra-cold atoms. 
}

\begin{document}

\maketitle

\section{Introduction}
Quantum engineering, i.e. coherent controlling of atomic systems, is a rapidly developing field of modern physics. Its development is important because it gives us the possibility of a deep study of quantum mechanics. 
Present experimental methods in quantum engineering allow one to prepare and control the system in a state with a rigorously chosen number of interacting atoms. Spectacular experiments on ultra-cold few-body fermionic systems confined in a one-dimensional harmonic trap \cite{Joachim0,Joachim1,Joachim2,Joachim3} have opened an alternative way to study strongly correlated quantum systems. It is believed that these kind of studies can shed a new light on our understanding of the crossover from the few-body world to the situation with a macroscopic number of quantum particles \cite{Joachim3}. They can help to answer the fundamental question on emerging of collective behavior from the two-body interactions. In particular it is not obvious what remains from the Cooper-pair correlations in a small system of interacting fermions.    

Recently, harmonically trapped effectively one dimensional few-body systems were intensively studied on theoretical \cite{Guan,Guan2,Cui,Volo,Blume,Sowinski,Zinner,Vicari,Santos,Amico2,Blume2,Mehta,Harshman} and experimental \cite{Joachim0,Joachim1,Joachim2,Joachim3} level. For example, it is known that in a strictly one-dimensional geometry, in the limit of infinite repulsion (known as Tonks-Girardeau limit) the bosonic system can be mapped on the system of noninteracting fermions \cite{Tonks,Girardeau,Lieb,Paredes}. Moreover, in this limit, the ground-state of the one-dimensional system of a few spin-1/2 fermions is highly degenerated \cite{Guan2,Sowinski} and a combination which is fully antisymmetric with respect to the position of fermions is identical to the ground state of the system of spinless fermions \cite{Girardeau2,Guan}. This degeneracy of the ground-state and its consequences for the experimental results were also studied \cite{Sowinski}. The  mapping mentioned above, in the limit of infinite repulsion was probed experimentally for a two-fermion system \cite{Joachim1}, for which the exact solution is known \cite{Busch,Esslinger,Gajda,Rontani,Idziaszek,Busch2}.

In a recent paper \cite{Joachim2} the attractive system was probed experimentally and formation of fermionic pairs was observed. This experiment opened an alternative path to the experimental exploration of the few-body systems. Therefore the quantitative predictions based on the first-principle theory of a few-body system are necessary. We should mentioned here that pairing in a system of fermionic atoms in the mean-filed Hartree-Fock approach was studied in \cite{Karpiuk,Swistak} where the energy gap separating ground and excited states was studied. This approach was recently exploited via exact diagonalization of a few-body Hamiltonian \cite{Massimo}. 

\section{The Model} In this Letter we study a few fermions of mass $m$, confined in a one-dimensional harmonic trap of frequency $\Omega$, mutually interacting via short-range delta-like attractive potential. In the second-quantization formalism the Hamiltonian of the model has a form:
\begin{align} \label{Hamiltonian}
\hat{\cal H} &=\sum_{\sigma\in\{\downarrow,\uparrow\}}\int\!\mathrm{d}x\,\hat\Psi_\sigma^\dagger(x)\left[-\frac{1}{2}\frac{\mathrm{d}^2}{\mathrm{d}x^2} + \frac{1}{2}x^2\right]\hat\Psi_\sigma(x) \nonumber \\
&+ g\int\!\mathrm{d}x\, \hat\Psi_\downarrow^\dagger(x)\hat\Psi_\uparrow^\dagger(x)\hat\Psi_\uparrow(x)\hat\Psi_\downarrow(x),
\end{align}
where $\Psi_\sigma(x)$ annihilates a fermion with spin $\sigma$ at a point $x$, and $g$ measures a strength of the contact interactions. For convenience we expressed all quantities in harmonic oscillator units, i.e. energies in $\hbar\Omega$, lengths in $\sqrt{\hbar/m\Omega}$, temperatures in $\hbar\Omega/k_B$ etc. The dimensionless interaction constant $g=(m/\hbar^3\Omega)^{1/2}g_{1D}$ is proportional to the effective one-dimensional interaction $g_{1D}$ between fermions of opposite spins. The Hamiltonian \eqref{Hamiltonian} commutes  separately with the number of fermions in each spin component  $\hat{N}_\sigma = \int\!\mathrm{d}x\,\Psi^\dagger_\sigma(x)\Psi_\sigma(x)$. Due to the fact that in the Hamiltonian there are no terms changing number of fermions in each spin component $N_\sigma$, particles can be treated as distinguishable. There are no symmetry conditions related to the exchange of particles with different spins and no physical results depend on the choice of commutators vs. anticommutators for opposite spins field operators (see \cite{Weinberg}). Here we assume that the field operators of different spins do commute, $\left[\hat\Psi_\uparrow(x),\hat\Psi^\dagger_\downarrow(x')\right] = \left[\hat\Psi_\uparrow(x),\hat\Psi_\downarrow(x')\right]=0$.   However, particles of the same spin satisfy the standard anti-commutation relations $\left\{\hat\Psi_\sigma(x),\hat\Psi^\dagger_\sigma(x')\right\}=\delta(x-x')$ and $\left\{\hat\Psi_\sigma(x),\hat\Psi_\sigma(x')\right\}=0$. We focus our attention on correlations between fermions with opposite spins in a balanced system of $N_\uparrow = N_\downarrow=N/2$. 
 
\section{The Method}
First, we fix the number of fermions in each component,  then we decompose the field operator $\Psi_\sigma(x)$ in the single-particle basis $\phi_n(x)$ of harmonic oscillator eigenstates with sufficiently large cutoff, $n\leq n_{max}$,  and finally, we diagonalize the full many-body Hamiltonian \eqref{Hamiltonian} in this basis via Implicitly Restarted Arnoldi Method \cite{Arnoldi}. In this way we find ${\cal N}$  eigenstates $|\mathtt{G}_i\rangle$ and the corresponding lowest eigenenergies $E_i$. In our case ${\cal N}=60$. 

\begin{figure}
\includegraphics{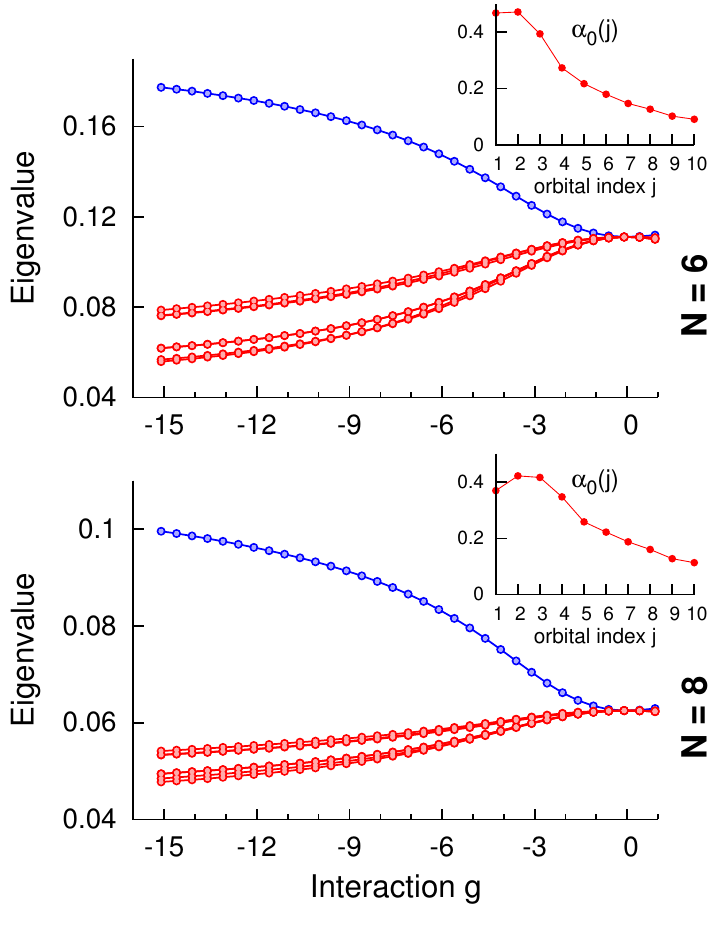}
\caption{First eight eigenvalues $\lambda_i$ of the reduced two-fermion density matrix $\rho^{(2)}$ as a function of interaction $g$, calculated in the ground-state of the system obtained numerically for $N_\uparrow=N_\downarrow=3$ and $4$. For attractive interactions one of the eigenvalues starts to dominate in the system indicating occurrence of the fermionic pair condensation.  In insets we show probability amplitudes $\alpha_0(j)$ for the dominant orbital of the correlated pair for $g=-10$ (see the main text).  \label{Fig1}}
\end{figure} 

\section{Properties of the ground state}
To get some intuition, first we discuss the properties of the system at zero temperature, i.e. we analyze the ground-state  $|\mathtt{G}_0\rangle$. In the limit of vanishing interaction $g$, the ground state of the system is simply formed by filling up the lowest single-particle states of the harmonic potential by pairs of opposite spin fermions, up to the Fermi level $n_F=N_\uparrow = N_\downarrow=N/2$. This fact is directly manifested in the spectral decomposition of the reduced two-fermion density matrix
\begin{multline}
\rho^{(2)}(x_1,x_2;x_1',x_2') =\\ \frac{4}{N^2}\times\langle \mathtt{G}_0|\hat{\Psi}^\dagger_\downarrow(x_1)\hat{\Psi}^\dagger_\uparrow(x_2)\hat{\Psi}_\uparrow(x_2')\hat{\Psi}_\downarrow(x_1')|\mathtt{G}_0\rangle.
\end{multline}
Let us notice that coordinates $x_1$ and $x_1'$ ($x_2$ and $x_2'$) correspond to the spin-down (spin-up)  fermions. From now on we will use this convention throughout our Letter. Distinction between the two species is possible because in our Hamiltonian there are no terms responsible for spin flipping mixing different species.  

\begin{figure}
\includegraphics{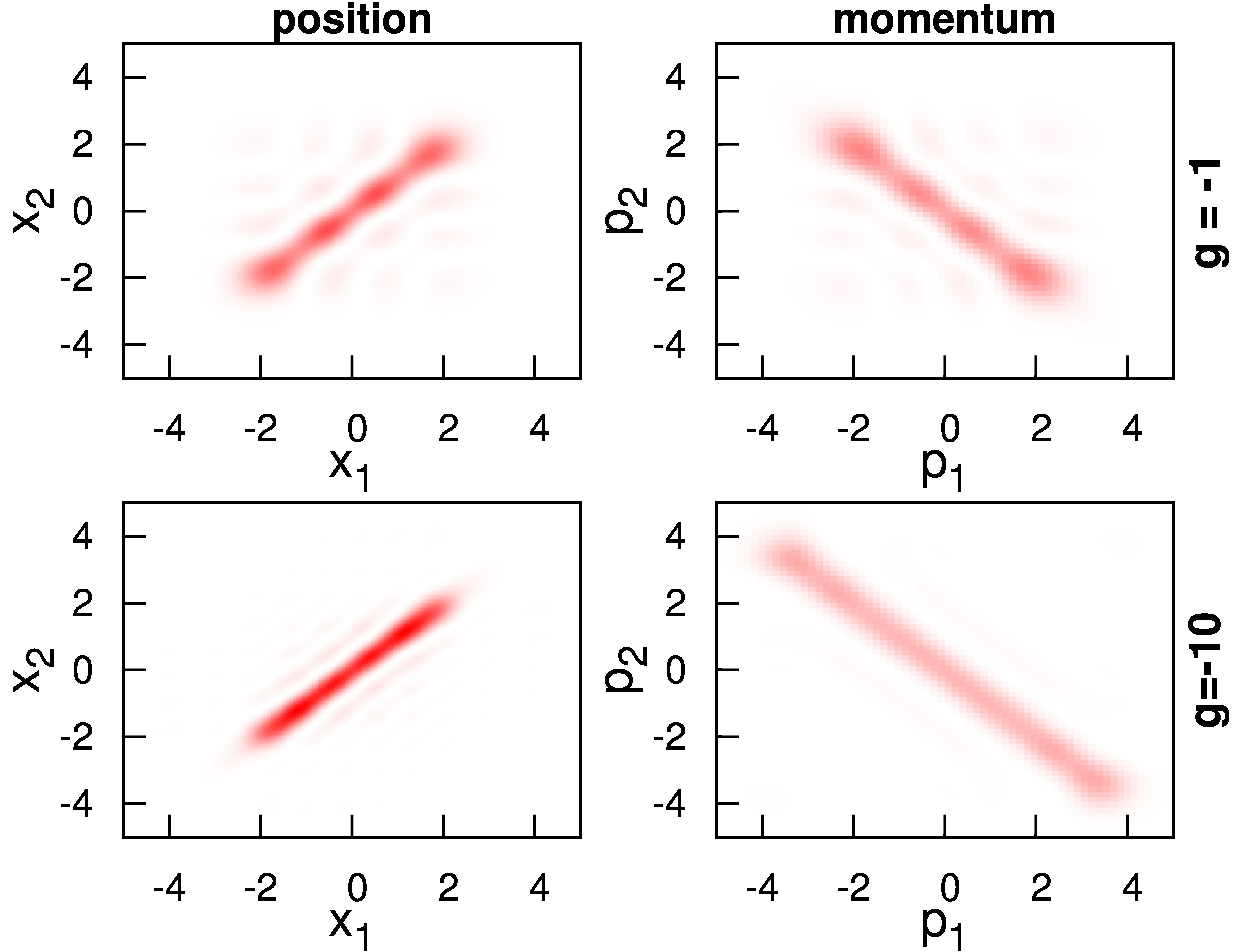}
\caption{The density distribution of finding fermions with opposite spins in the dominant orbital in the position $|\psi_0(x_1,x_2)|^2$ (left panels) and in the momentum domain $|\psi_0(p_1,p_2)|^2$ (right panels). Strong correlations in positions and momenta are enhanced for stronger attractions.  \label{Fig2}}
\end{figure}

All nonzero eigenvalues of $\rho^{(2)}$ are equal to $\lambda =4/N^2$. They determine the relative populations of the corresponding eigenvectors, i.e. the two-particle orbitals $\psi_i(x_1,x_2)$ describing possible pairings of fermions with opposite spins. Situation changes dramatically when attractive interaction is switched on. Then, we observe that one eigenvalue starts to dominate in the system (Fig. \ref{Fig1}). According to the Penrose-Onsager criterion, an occurrence of the dominant orbital in the spectral decomposition of the reduced density matrix is an indication of the condensation of fermionic pairs in the selected orbital. For convenience we  denote the dominant orbital and its eigenvalue by $\psi_0$ and $\lambda_0$ respectively. To get a better understanding of the structure of the ground state of the system it is convenient to relate the orbitals $\psi_i(x_1,x_2)$ to the orbitals $\varphi_i(x)$ -- the eigenvectors of the reduced single-fermion density matrices
\begin{equation}
\rho^{(1)}(x;x') = \frac{2}{N}\times \langle\mathtt{G}_0|\hat{\Psi}^\dagger_\sigma(x)\hat{\Psi}_\sigma(x')|\mathtt{G}_0\rangle,
\end{equation}
where $\sigma\in\{\uparrow,\downarrow\}$. Since we assumed the exact symmetry between `up' and `down' fermions, $\rho^{(1)}$ is exactly the same for both spin components. By projecting two-fermion orbitals of $\rho^{(2)}$ on the product of the single-fermion orbitals of $\rho^{(1)}$ we find that the set of all orbitals $\psi_i(x_1,x_2)$ can be divided into two classes.  For a given two-fermion orbital $\psi_i$ belonging to the first class one can find exactly two single-fermion orbitals $\varphi_j$ and $\varphi_k$ ($j \neq k$) such that 
\begin{subequations}
\begin{equation}
\label{eq.4}
\psi_i(x_1,x_2) = \left[\varphi_j(x_1)\varphi_k(x_2) \pm \varphi_k(x_1)\varphi_j(x_2)\right]/\sqrt{2}.
\end{equation}
On the contrary, in the second class each two-fermion orbital $\psi_i(x_1,x_2)$ has a structure of the correlated pair -- the superposition of states with two particles of opposite spin occupying the same single-particle state:
\begin{equation}\label{CooperFunction}
\psi_i(x_1,x_2)=\sum_j \alpha_i(j)\, \varphi_j(x_1)\varphi_j(x_2),
\end{equation}
\end{subequations}
where $\sum_j|\alpha_i(j)|^2=1$.
From our numerical analysis it follows that the dominant orbital, $\psi_0(x_1,x_2)$, always belongs to the second class and is symmetric with respect to exchange of spin-up and spin-down fermions.
Moreover,   the expectation value of the operator 
\begin{equation}\label{PairAnihilator}
\hat{\cal A}=\sum_{i,j,k} \hat{b}_{\downarrow i}^\dagger \hat{b}_{\uparrow j}^\dagger \hat{b}_{\uparrow k}\hat{b}_{\downarrow k},
\end{equation}
vanishes in this orbital only. The operator $\hat{b}_{\sigma i}$ annihilates a fermion in the orbital $\varphi_i$ of the reduced single-fermion density matrix $\rho^{(1)}$. 
Expectation value ${\cal S}_0 = \langle{\mathtt G}_0| \hat{\cal A} |{\mathtt G}_0\rangle$ is a direct counterpart of the superconducting order parameter ${\cal S}=\int\mathrm{d}\boldsymbol{r}\langle \hat\Psi_\uparrow(\boldsymbol{r})\hat\Psi_\downarrow(\boldsymbol{r})\rangle$ known from the BCS theory \cite{BCS}.  Operator $\hat{\cal A}$ can be viewed as the number-conserving  destruction operator of the correlated pair.  

The strong correlation in the dominant orbital can be observed in the probability density of finding fermions with opposite spins (Fig. \ref{Fig2}). The probability distribution is located mainly along the diagonal $x_1 =x_2$. The dispersion of  $(x_1-x_2)$ is a measure of  a size of the correlated pair which can be compared to the size of the whole many body system.  The latter is not the single particle extension  determined by the characteristic length of the ground state of the external trap \cite{Gajda1}, in  particular for attractive systems. The size of the system can be determined from a correlated many-body detection  \cite{Gajda2}. To this end we define the correlated two-fermion observables: (i)  for the whole system and (ii) for the dominant orbital. In both cases they are related to the conditional probability distribution: the probability of a particle to be at a position $x$ provided that the second one is at $x_0$ \cite{Gajda2}
\begin{equation} \label{condprob}
{\cal P}(x|x_0) = \frac{{\cal P}(x,x_0)}{{\cal P}(x_0)}.
\end{equation}
Probabilities ${\cal P}(x,x_0)$ (${\cal P}(x_0)$) are related to the two-particle (one-particle) density matrix of the whole system, or the two-particle (one-particle) density of the dominant orbital, respectively. 

\begin{figure}
\includegraphics{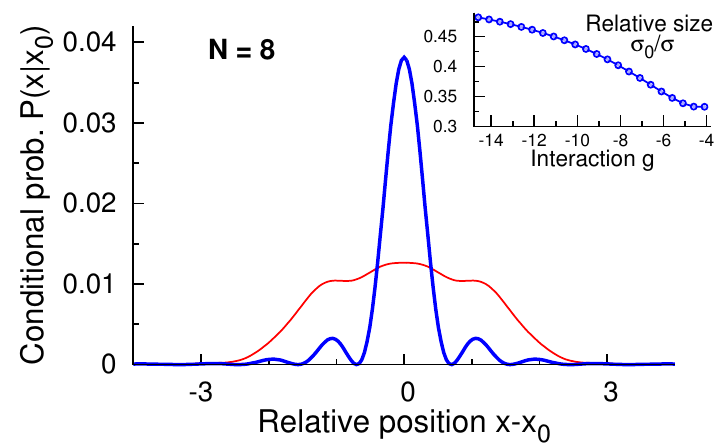}
\caption{Conditional probability distributions \eqref{condprob} of finding a particle at position $x$ provided that the second one is located at $x_0=0$ for $g=-12$. The thin red and thick blue lines correspond to the whole system and dominant orbital respectively. Note that although they look different their spreadings are comparable. Inset: Relative size of the dominant orbital $\sigma_0/\sigma$ as a function of the interaction constant $g$. Note, that the size of the correlated pair is only about two times smaller than the size of the system. \label{Fig4}}
\end{figure}

In Fig. \ref{Fig4} we show an example of these distributions for $g=-12$ and $x_0=0$. The probability distribution for the dominant orbital, contrary to the probability of the whole system, oscillates.  However, the spatial extensions of both these distributions are similar. We can quantify this by comparing he size of the dominant orbital and the size of the entire system. They can be defined as the dispersions $\sigma$ and $\sigma_0$ of the corresponding probability distributions averaged over the position $x_0$. For example the size of the system is
\begin{equation}
\sigma = \left(\int\!\!\int\!\!\mathrm{d}x\,\mathrm{d}x_0 (x-x_0)^2\,{\cal P}(x|x_0){\cal P}(x_0)\right)^{1/2}.
\end{equation}
In the inset of Fig. \ref{Fig4} we show the ratio of $\sigma_0/\sigma$ as a function of the interaction constant $g$. The average size of the correlated pair is about two up to three times smaller than the size of the system for strong attraction. This holds in the whole range of interactions studied. 

In the momentum space we find some anti-correlation between momenta of two fermions $p_1$ and $p_2$ occupying the dominant orbital (Fig. \ref{Fig2}). Note, that the probability distribution is located mainly along the anti-diagonal and has finite and relatively large ($\sim \hbar/\sigma$) spreading. This is not surprising since for small confined systems the exact anti-correlation in the momentum space is not expected \cite{Bruun1,Bruun2,Bruun3}. 

The dominant orbital $|\psi_0\rangle=\sum_j\alpha_0(j)\, \hat{b}^\dagger_{\downarrow j} \hat{b}^\dagger_{\uparrow j} |\Omega\rangle$ is an analog of the Cooper pair \cite{Cooper}. The two atoms in the Cooper pair occupy simultaneously the same one-particle orbital $\varphi_j$ with the probability $|\alpha_0(j)|^2$. As indicated  in insets in Fig.\ref{Fig1} the probabilities are large for $j=1,\ldots, N/2$. The corresponding orbitals have large and similar occupations, however with increasing number of atoms a small maximum appears. Evidently the wavefunction of pairs found here is a superposition of products of all occupied single particle orbitals -- not only of the orbitals from the Fermi surface as in the case of the standard Cooper pairs.

\begin{figure}
\includegraphics{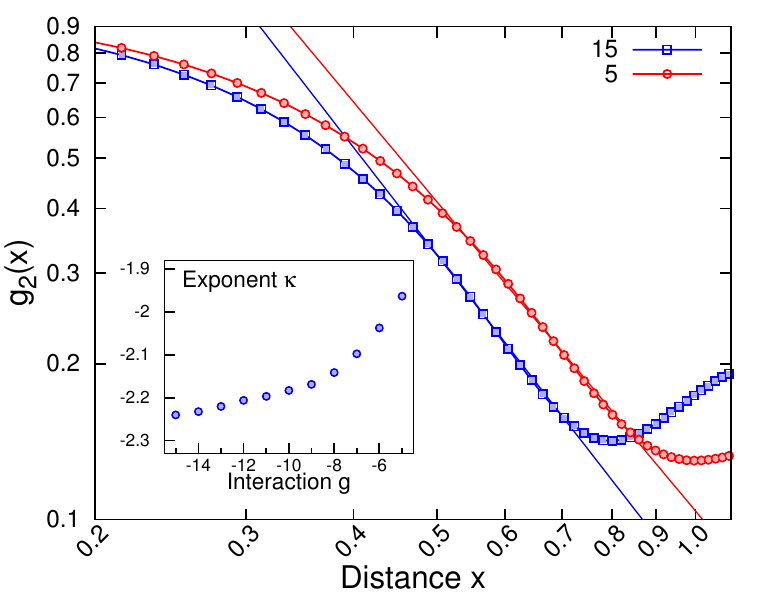}
\caption{Off-diagonal correlation $g_2(x)$ for interaction parameter $g=-5$ (red circles) and $g=-15$ (blue squares). The lines indicate a power law decay at distance comparable to the size of the Cooper-like pair. The exponent of the power law decay as a function of the interaction parameter $g$ is shown in the inset. \label{Fig5}}
\end{figure}

The superfluid system of Cooper pairs is often viewed as  the Bose-Einstein condensate of bosonic particles -- the fermionic pairs. Such a system should show the off-diagonal long range order. In the case of a small trapped system in 1D, this order results in a power law decay of the off-diagonal elements of the second order correlation function at distances comparable to the system's size as shown in exact calculations for trapped bosons \cite{Schmidt}. In Fig.\ref{Fig5} we show the off-diagonal correlation parameter, $g_2(x)=\rho^{(2)}(0,0;x,x)/\sqrt{\rho^{(2)}(0,0;0,0)\rho^{(2)}(x,x;x,x)}$, as a function of the distance $x$. At small distances the off-diagonal correlation is large and practically constant indicating a two-particle coherence. However, at distances comparable to the size of the correlated pair (compare Fig.\ref{Fig4}) the off-diagonal correlation falls down as expected, i.e. according to the power law $g_2(x)\sim x^\kappa$. Exponent $\kappa$ depends on the interaction constant $g$ (inset of Fig.\ref{Fig5}). The dependence on interactions is rather weak.  This can be understood by noticing a small contribution of the Cooper-like pairs to the two-particle density -- the system studied here corresponds to a Bose-Einstein condensate with a very large quantum depletion. This is also the reason why the power law decay of $g_2(x)$ is not observed for $g\gtrsim -5$.

All our observations  support the point of view that the system studied here reminds in many respects  the superfluid system of the Cooper-like pairs with all the differences originating from the fact that our system is 1D, nonuniform and microscopic. However, it is not the system of weakly bound fermionic pairs near a Feshbach resonance as observed in the seminal experiments related to the Bose-Einstein condensation of fermionic pairs in the  BEC-BCS  regime \cite{Jin,Ketterle,Thomas,Salomon}. Theoretical description of the condensation of Feshbach molecules requires taking into account the presence of the Feshbach resonance and is out of the scope of our study.

\section{Non zero temperatures} Now, we discuss the situation in which the system studied is in contact with a thermostat of given temperature $T$. Therefore, instead of the many-body ground state $|\mathtt{G}_0\rangle$, we assume that the state of the system is well described by the density matrix operator $\hat\rho_{\mathtt T} = {\cal Z}^{-1}\exp(-\hat{\cal H}/T)$, where ${\cal Z}$ is a canonical partition function assuring  $\mathrm{Tr}\,\hat\rho_{\mathtt T} = 1$. Numerically, the density matrix operator $\hat\rho_{\mathtt T}$ is constructed from a finite number of ${\cal N}=60$ eigenstates $|\mathtt{G}_i\rangle$ of the full many-body Hamiltonian as follows $\hat\rho_{\mathtt T}=\sum_i p_i |\mathtt{G}_i\rangle\langle\mathtt{G}_i|$, where $p_i = \exp(-E_i/T)/\sum_i{\exp(-E_i/T)}$. In analogy to the zero temperature case, we define a reduced two-fermion density matrix 
\begin{multline} \label{2FCorelTemp}
\rho^{(2)}_{\mathtt{T}}(x_1,x_1';x_2,x_2') = \\ \frac{4}{N^2}\times \mathrm{Tr}\left[\hat{\rho}_{\mathtt{T}}\,\hat{\Psi}^\dagger_\downarrow(x_1)\hat{\Psi}^\dagger_\uparrow(x_2)\hat{\Psi}_\uparrow(x_2')\hat{\Psi}_\downarrow(x_1')\right]
\end{multline}
and we diagonalize it. We perform the whole spectral analysis of $\rho^{(2)}_{\mathtt T}$ in analogy to the zero temperature limit. First, we find temperature dependent single-particle orbitals annihilated by temperature dependent operators $\hat b_{\sigma i}$. These states serve as a basis for two-fermion orbitals of $\rho^{(2)}_{\mathtt{T}}$. We find that the division of all orbitals into two classes with respect to their representation by finite-temperature single-fermion orbitals holds. In particular coefficients $\alpha_0(j)$ practically do not depend on the temperature. We also checked that the dominant orbital of $\rho^{(2)}_{\mathtt T}$ remains the only orbital with non-vanishing expectation value of the particle conserving destruction operator $\hat{\cal A}$ (defined with temperature dependent $\hat b_{\sigma i}$). This means that the superconducting order parameter ${\cal S}(T)=\mathrm{Tr}\left[\hat\rho^{(2)}_\mathtt{T}\hat{\cal A}\right]$ is well defined. Exploring the fact that coefficients $\alpha_0(j)$ do not depend on temperature, the order parameter ${\cal S}(T)$ can be directly related to the superconducting fraction represented by the dominant eigenvalue at given temperature $\lambda_0(T)$:
\begin{equation}
{\cal S}=\frac{\lambda_0(T)}{\lambda_0(0)}{\cal S}_0.
\end{equation} 

\begin{figure}
\includegraphics{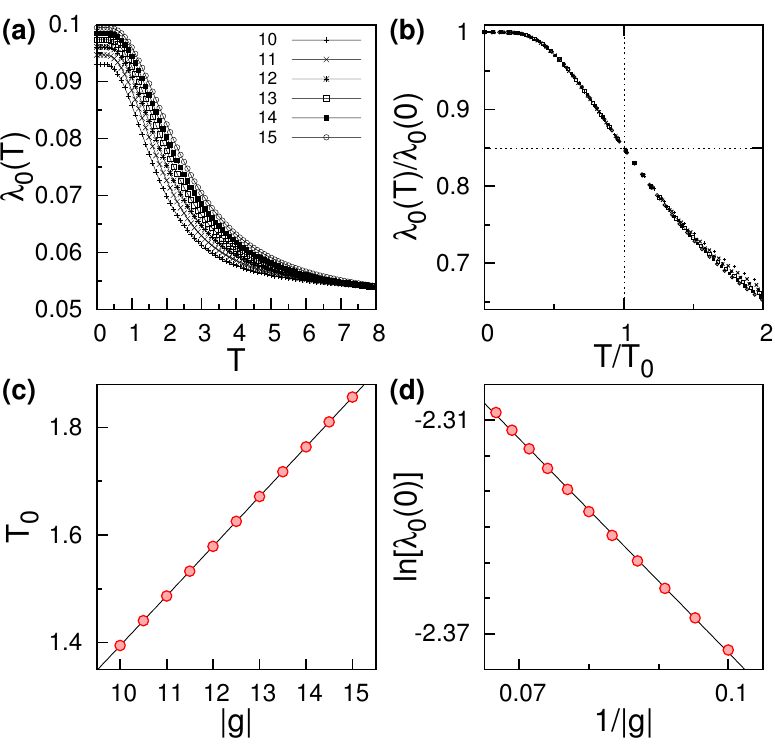}
\caption{Properties of the superconducting fraction $\lambda_0(T)$ for $N_\uparrow=N_\downarrow=4$. (a) Dominant eigenvalue $\lambda_0(T)$ of the two-fermion density matrix $\rho_\mathtt{T}^{(2)}$ as a function of temperature for different two-body interaction constant $g$. (b) Rescaled $\lambda_0(T)/\lambda_0(0)$ as a function of rescaled temperature $T/T_0$ for different two-body interaction constant $g$. All numerical points collapse to the one universal curve, which confirms the universal behavior of $\lambda_0(T)$. (c) Characteristic temperature $T_0$ as a function of interaction. Numerical data points fit almost perfectly to the linear regression. (d) Zero temperature superconducting fraction $\lambda_0(0)$ as a function of interaction. In chosen scaling numerical data points fit to the linear regression, which confirms universal behavior \eqref{LambdaReg}. \label{Fig6}}
\end{figure}

With increasing temperature we observe a rapid decay of the occupation of the dominant orbital (Fig. \ref{Fig6}a) and as consequence, the superconducting fraction tends to zero. It is worth to notice that the dominant eigenvalue $\lambda_0(T)$ of the reduced two-fermion density matrix for the system studied is closely related to the gap function $\Delta$ known from the theory of superconductivity for isotropic system of interacting fermions in the thermodynamic limit. 
Therefore it is quite obvious that the behavior of $\lambda_0(T)$ is in one-to-one correspondence with the behavior of the gap function $\Delta$ known from the BCS theory.  

In Fig. \ref{Fig6}a we plot the temperature dependence of the superconducting fraction $\lambda_0(T)$ for $N_\downarrow=N_\uparrow=4$ and different interactions $g$. As can be seen, this quantity  crucially depends on the interaction constant $g$. Nevertheless, it has some fundamental universality which indicates correspondence to the BCS theory. To show this universality we define two natural quantities which depend on $g$, i.e. superconducting fraction at zero temperature, $\lambda_0(0)$, and the characteristic temperature $T_0$ at which, by  definition, the superconducting fraction drops by some chosen factor $\eta={\lambda_0}(T_0)/{\lambda_0}(0)$. We checked that our predictions are  insensitive to any reasonable choice of the value of $\eta$ and in the following we set $\eta=85\%$. Then, by plotting the normalized data, i.e.  $\lambda_0(T)/\lambda_0(0)$ vs. $T/T_0$, we find that all numerical  points collapse to the  universal curve which does not depend on details of the model (see Fig.\ref{Fig6}b). We checked that this universal behavior is independent of the definition of the characteristic temperature $T_0$. Moreover, we find that the characteristic temperature $T_0$ is a linear function of interaction $g$ (Fig. \ref{Fig6}c). On the other hand, the dependence of the zero temperature superconducting fraction $\lambda_0(0)$ on the interaction constant $g$ is exactly the same as the dependence of the gap-function in the BCS theory (Fig. \ref{Fig6}d)
\begin{equation} \label{LambdaReg}
\lambda_0(0) = \Lambda\mathrm{e}^{-1/{\cal M}|g|}.
\end{equation}
Amazingly, the value of the parameter $\cal M$ obtained from the numerical fit to the data is equal to $2.001\pm0.013$, which is in a perfect agreement with the intuitive picture of the theory of superconductivity. In the case of homogenous and infinite system, the theory predicts that $\cal M$ should be equal to the density of states at the Fermi level which in our one-dimensional case is equal to 2. The result, seemingly clear, is not so obvious. After all, the pairs found here are composed not only of particles from the Fermi sea. Finally, let us notice that dependence of the characteristic temperature $T_0$ on $g$ is linear which is different with respect to the BCS theory for a homogeneous and 3D system. This comes from the fact that the number of particles here is small and the system is one-dimensional.   

\begin{figure}
\includegraphics{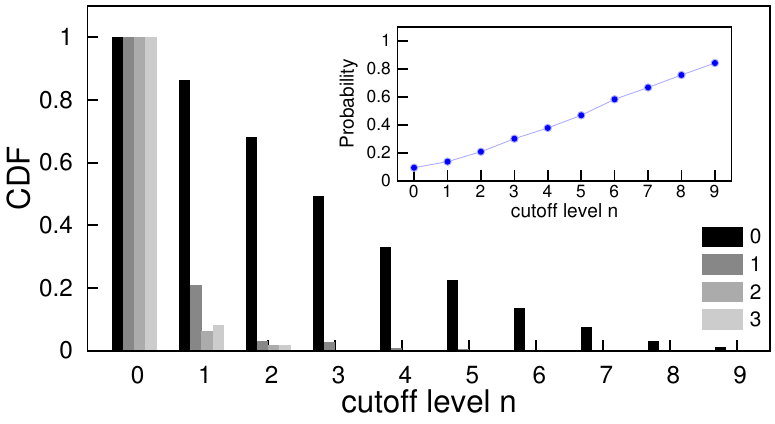}
\caption{Two-particle cumulative distribution function (CDF) for few first orbitals $\psi_i$ at $T=0$ and $g=-10$ for $N=8$ (the gray scale corresponds to the orbital index $i$ as indicated in the legend). The hight of the peak is equal to the probability of finding simultaneously two particles (described by $\psi_i$) above chosen harmonic oscillator level $n$. In the inset we show the probability that the pair detected above a level $n$ comes from the dominant strongly correlated orbital $\psi_0$. Note, that for large $n$ the pair, if detected, is the Cooper-like pair with probability approaching one. \label{Fig7}}
\end{figure}

\section{Final remarks} We believe that our predictions can be verified in current experiments with ultra-cold fermions in optical traps. By lowering the edge of the trap potential (similarly to \cite{Joachim0}) the particles of energies above selected harmonic oscillator level are allowed to escape and can be detected. The probability of simultaneous detection of two particles above selected energy $n$ is related to the two-particle cumulative distribution function ${\cal C}(n)$ (CDF). This cumulative distribution we calculate for every orbital $\psi_i$  of the two-fermion density matrix $\rho^{(2)}$ separately by summing the probabilities of finding  simultaneously the two fermions on the harmonic level with index not smaller then $n$. In Fig. \ref{Fig7} the CDF for a few first orbitals $\psi_i$  of the two-fermion density matrix $\rho^{(2)}$ ($i=0,\ldots,3$) at $T=0$ and $g=-10$ for $N=8$ is shown. The gray scale corresponds to the orbital index $i$ as indicated in the legend. The hight of the peak is equal to the probability of finding simultaneously two particles (described by $\psi_i$) above chosen harmonic oscillator level $n$. In the inset we show the probability that the pair detected above a level $n$ comes from the Cooper-like orbital $\psi_0$. Note that for large $n$, the pair, if detected, is the Cooper-like pair with probability approaching one. In this case, due to the correlations described above, both particles should be preferably detected with opposite momenta.  Other pairs have very low probability to be found above the energy $n$.
The figure clearly shows that if the pair is detected  above the harmonic oscillator level $n=9$ (both particles must have energies larger than $n$) then with probability of about $85\%$ it has to be the correlated Cooper-like pair initially occupying the dominant orbital. In this way the correlated pair can be distilled from the trap. Measuring the two particle correlation function of the pair in the momentum domain  (Fig. \ref{Fig2}) is the ultimate proof of detection of the Cooper pair. We checked that this is true also at finite temperatures.  

\section{Acnowledgements} This research was supported by the (Polish) National Science Center Grant No. DEC-2012/04/A/ST2/00090.


\begin{thebibliography}{99}
\bibitem{Joachim0}
  \Name{Serwane F., Z\"urn G., Lompe T., Ottenstein T. B., Wenz A. N., \and Jochim S.}
  \REVIEW{Science}{332}{2011}{336}.
\bibitem{Joachim1} 
  \Name{Z\"urn G. \etal}
  \REVIEW{Phys. Rev. Lett.}{108}{2012}{075303}.
\bibitem{Joachim2}
  \Name{Z\"urn G. \etal}
  \REVIEW{Phys. Rev. Lett.}{\bf 111}{2013}{175302}.
\bibitem{Joachim3}
  \Name{Wenz A. N. \etal}
  \REVIEW{Science}{342}{2013}{457}.
\bibitem{Guan}
  \Name{Guan L. \and Chen S.}
  \REVIEW{Phys. Rev. Lett.}{105}{2010}{175301}.
\bibitem{Guan2}
  \Name{Guan L., Chen S., Wang Y., \and Ma Z.-Q.}
  \REVIEW{Phys. Rev. Lett.}{102}{2009}{160402}.
\bibitem{Cui} 
  \Name{Cui X. \and Ho T.-L.}
  \REVIEW{Phys. Rev. A}{89}{2014}{023611}.
\bibitem{Volo}
  \Name{Volosniev A. G. \etal}
  \Review{Few-Body Syst.}{55}{2014}{839}.
\bibitem{Blume}
  \Name{Gharashi S. E. \and Blume D.}
  \REVIEW{Phys. Rev. Lett.}{111}{2013}{045302}.
\bibitem{Sowinski}
  \Name{Sowi\'nski T., T. Grass, O. Dutta, \and M. Lewenstein}
  \REVIEW{Phys. Rev. A}{88}{2013}{033607}.
\bibitem{Zinner}
  \Name{Lindgren E. J. \etal}
  \REVIEW{New J. Phys.}{16}{2014}{063003}.
\bibitem{Vicari}
  \Name{Angelone A., Campostrini M., Vicari E.}
  \REVIEW{Phys. Rev. A}{89}{2014}{023635}.
\bibitem{Santos}
  \Name{Deuretzbacher F. \etal}
  \REVIEW{Phys. Rev. A}{90}{2014}{013611}.
\bibitem{Amico2}
  \Name{D’Amico P. \and Rontani M.}
  \REVIEW{J. Phys. B}{47}{2014}{065303}.
\bibitem{Blume2}
  \Name{Yan Y. Q. \and Blume D.}
  \REVIEW{Phys. Rev. A}{90}{2014}{013620}.
\bibitem{Mehta}
  \Name{Mehta N. P.}
  \REVIEW{Phys. Rev. A}{89}{2014}{052706}.
\bibitem{Harshman}
  \Name{Harshman N. L.}
  \REVIEW{Phys. Rev. A}{89}{2014}{033633}.
\bibitem{Tonks}
  \Name{Tonks L.}
  \REVIEW{Phys. Rev.}{50}{1936}{955}.
\bibitem{Girardeau}
  \Name{Girardeau M.}
  \REVIEW{J. Math. Phys.}{1}{1960}{516}.
\bibitem{Lieb}
  \Name{Lieb E. \and Liniger W.}
  \REVIEW{Phys. Rev.}{130}{1963}{1605}.
\bibitem{Paredes}
  \Name{Paredes B. \etal}
  \REVIEW{Nature (London)}{429}{2004}{277}.
\bibitem{Girardeau2}
  \Name{Girardeau M. D.}
  \REVIEW{Phys. Rev. A}{82}{2010}{011607(R)}.
\bibitem{Busch}
  \Name{Busch T., Englert B.-G., Rz\c a\.zewski K., \and Wilkens M.}
  \REVIEW{Found. Phys.}{28}{1998}{549}.
\bibitem{Gajda}
  \Name{Sowi\'nski T., Brewczyk M., Gajda M., \and Rz\c a\.zewski K.}
  \REVIEW{Phys. Rev. A}{82}{053631}{2010}.
\bibitem{Esslinger}
  \Name{Michael K\"ohl \etal}
  \REVIEW{J. Phys. B: At. Mol. Opt. Phys.}{39}{2006}{S47}.
\bibitem{Rontani}
  \Name{Rontani M.}
  \REVIEW{Phys. Rev. Lett.}{108}{2012}{115302}.
\bibitem{Idziaszek}
  \Name{Idziaszek Z. \and Calarco T.}
  \REVIEW{Phys. Rev. A}{74}{2006}{022712}.
\bibitem{Busch2}
  \Name{Campbell S. \etal}
  \REVIEW{Phys. Rev. A}{90}{2014}{013617}.
\bibitem{Karpiuk}
  \Name{Karpiuk T., Brewczyk M., \and Rz\c a\.zewski K.}
  \REVIEW{Phys. Rev. A}{69}{2004}{043603}.  
\bibitem{Swistak}
  \Name{\'Swis\l ocki T., Karpiuk T., \and Brewczyk M.}
  \REVIEW{Phys. Rev. A}{77}{2008}{033603}. 
\bibitem{Massimo}
  \Name{D’Amico P. \and Rontani M.}
  {ArXiv:1404.7762 preprint, 2014}.
\bibitem{Weinberg}
  \Name{Weinberg S.}
  \Book{The Quantum Theory of Fields}
  \Vol{1}
  \Publ{Cambridge University Press, Cambridge}
  \Year{1995}.
\bibitem{Arnoldi}
  \Name{Lehoucq R. B., Sorensen D. C., \and Yang C.}
  \Book{Arpack User's Guide: Solution of Large-Scale Eigenvalue Problems With Implicityly Restorted Arnoldi Methods}
  \Publ{Society for Industrial \& Applied Mathematics}
  \Year{1998}.
\bibitem{BCS}
  \Name{Bardeen J., Cooper L. N., Schrieffer J. R.}
  \REVIEW{Phys. Rev.}{108}{1957}{1175}.
\bibitem{Gajda1}
  \Name{Gajda M., Za\l uska-Kotur M., \and Mostowski J.}
  \REVIEW{J. Phys. B: At. Mol. and Opt. Phys.}{33}{2000}{4003}. 
\bibitem{Gajda2}
  \Name{Gajda M.}
  \REVIEW{Phys Rev. A}{73}{2006}{023603}.
\bibitem{Bruun1}
  \Name{Bruun G. M. \and Heiselberg H.}
  \REVIEW{Phys. Rev. A}{65}{2007}{053407}.
\bibitem{Bruun2}
  \Name{Bruun G. M.}
  \REVIEW{Phys Rev A}{66}{2002}{041602}.
\bibitem{Bruun3}
  \Name{Viverit L. \etal}
  \REVIEW{Phys. Rev. Lett.}{93}{2004}{110406}.
\bibitem{Cooper}
  \Name{Cooper L. N.}
  \REVIEW{Phys. Rev.}{104}{1956}{1189}.
\bibitem{Schmidt}
  \Name{Scmidt B. \and M. Fleischhauer}
  \REVIEW{Phys. Rev. A}{75}{2007}{021601(R)}.
\bibitem{Jin} 
  \Name{Regal C. A., Greiner M., \and Jin D. S.}
  \REVIEW{Phys. Rev. Lett.}{92}{2004}{040403}. 
\bibitem{Ketterle} 
  \Name{Zwierlein M. W. \etal}
  \REVIEW{Phys. Rev. Lett.}{92}{2004}{120403}.  
\bibitem{Thomas}
  \Name{Kinast J. \etal}
  \REVIEW{Phys. Rev. Lett.}{92}{2004}{150402}.
\bibitem{Salomon}
  \Name{Bourdel T. \etal}
  \REVIEW{Phys. Rev. Lett.}{93}{2004}{050401}.
\end{thebibliography}
\end{document}